\documentstyle[twoside,fleqn,espcrc2]{article}

\newcommand{\eqn}[1]{(\ref{#1})}
\newcommand{\ft}[2]{{\textstyle\frac{#1}{#2}}}
\def\be{\begin{equation}}
\def\ee{\end{equation}}
\def\bea{\begin{eqnarray}}
\def\eea{\end{eqnarray}}

\renewcommand{\a}{\alpha}

\renewcommand{\d}{\delta}
\newcommand{\pa}{\partial}

\newcommand{\m}{\mu}
\newcommand{\n}{\nu}

\def\cN{{\cal N}}

\def\IZ{{\hbox{{\rm Z}\kern-.4em\hbox{\rm Z}}}}

\def\bigone{{\hbox{1\kern -.23em{\rm l}}}}
\title{N=2 Electric-magnetic duality in a chiral background }
\author{Bernard de Wit\address{Institute for Theoretical Physics,
Utrecht University\\
Princetonplein 5, 3508 TA Utrecht, The Netherlands }}
\begin{document}
\setlength{\arraycolsep}{0pt}
\begin{titlepage}
\begin{flushright} THU-96/07\\ hep-th/9602060
\end{flushright}
\vfill
\begin{center}
{\large\bf N=2 Electric-magnetic duality in a chiral background${}^\dagger$}\\
\vskip 7.mm
{Bernard de Wit }\\
\vskip 0.1cm                                                      
{\em Institute for Theoretical Physics} \\
{\em Utrecht University}\\
{\em Princetonplein 5, 3508 TA Utrecht, The Netherlands} 
\end{center}
\vfill
\begin{center}
{\bf ABSTRACT}
\end{center}
\begin{quote}
We establish the consistency of duality transformations for 
generic systems of $N=2$ vector supermultiplets in the presence 
of a chiral background field. This is relevant, for instance, 
when dealing with spurion fields or when considering 
higher-derivative couplings of vector multiplets to supergravity. 
We point out that under duality most quantities do not transform 
as functions. With few exceptions, true functions are 
nonholomorphic, even though the duality transformations 
themselves are holomorphic in nature.  
\vfill      \hrule width 5.cm
\vskip 2.mm
{\small\small
\noindent $^\dagger$ Invited talk given at the $29^{\rm th}$ 
International Symposium Ahrenshoop on the Theory of Elementary 
Particles, Buckow, August 29 - September 2, 1995; to be published 
in the proceedings.}  
\end{quote}
\begin{flushleft}
February 1996
\end{flushleft}
\end{titlepage}
\begin{abstract}
We establish the consistency of duality transformations for 
generic systems of $N=2$ vector supermultiplets in the presence 
of a chiral background field. This is relevant, for instance, 
when dealing with spurion fields or when considering 
higher-derivative couplings of vector multiplets to supergravity. 
We point out that under duality most quantities do not transform 
as functions. With few exceptions, true functions are 
nonholomorphic, even though the duality transformations 
themselves are holomorphic in nature.  
\end{abstract}
\maketitle
\setcounter{footnote}{0}
\section{Introduction}
As is well known systems of abelian $N=2$ vector supermultiplets 
transform systematically under duality transformations: 
transformations that act on the (abelian) field strengths and 
rotate the combined field equations  
and Bianchi identities by means of a symplectic matrix. This was 
first exploited for pure $N=2$  
supergravity \cite{SG}. For generic $N=2$ vector supermultiplets it 
was discovered \cite{DWVP} that these 
transformations rotate the scalar fields $X^I$ and the derivatives 
$F_I$ of the holomorphic function $F(X)$ that encodes the 
Lagrangian, by means of an $Sp(2n+2;{\bf R})$ transformation, 
where $n$ denotes the number of vector multiplets\footnote{%
   Not counting the graviphoton of $N=2$ supergravity. In the 
   rigid case the symplectic matrix is only $2n$-dimensional. 
   Nonperturbatively, the symplectic tranformations are  
   usually restricted to a discrete subgroup.}. %
Initially the emphasis was on invariances of the equations of 
motion. The fact that the scalars in supergravity often 
parametrize an homogeneous  
space whose transitive isometries are realized by duality 
transformations, enables one to conveniently controll the 
nonpolynomial dependence on the scalar fields. Later it was 
realized that these transformations can also be used to 
reparametrize the theory in terms of a different function $\tilde 
F(\tilde X)$ \cite{CecFerGir}. For the subgroup of the 
symplectic group corresponding to an invariance of the equations 
of motion, the function remains the same. 

More recently symplectic reparametrizations were exploited by 
Seiberg and Witten  
\cite{SW} and later by others \cite{Lerche}, in 
obtaining exact solutions of low-energy effective actions for 
$N=2$ supersymmetric Yang-Mills theory. The 
singularities in these effective actions signal their breakdown 
due to the emergence of massless states corresponding to 
monopoles and dyons.  Although these states are the result of 
nonperturbative dynamcis, they are nevertheless accessable 
because at these points one conveniently converts to an 
alternative dual formulation, in which local field theory is 
again applicable. In many of the nonperturbative solutions the 
quantities $(X^I,F_J)$ can be 
defined as the periods of a meromorphic differential 
corresponding to a class of hyperelliptic curves. A similar 
phenonomenon is known from type-II string 
compactifications on Calabi-Yau manifolds, where $(X^I,F_J)$ can 
be associated with the periods of the $(3,0)$ form of the 
Calabi-Yau three-fold.

\section{$N=2$ Vector multiplets} 
The actions we use are based on $N=2$ chiral superspace 
integrals,
\be
S\propto {\rm Im}\;\Big(\int {\rm d}^4x \;{\rm d}^4\theta\; 
F(W^I)\Big)\,,  \label{chiral}
\ee
where $F$ is is an arbitrary function of reduced chiral multiplets 
$W^I(x,\theta)$. Such multiplets carry the gauge-covariant degrees of 
freedom of a vector multiplet, consisting of a complex scalar 
$X^I$, a spinor doublet $\Omega^{iI}$, a selfdual field-strength 
$F_{\m\n}^{-I}$ and a triplet of auxiliary fields $Y^I_{ij}$. 
This Lagrangian may coincide with the effective Lagrangian associated  
with some supersymmetric Yang-Mills theory, but for 
our purposes its origin is not directly relevant. To enable 
coupling to supergravity the holomorphic function should be 
homogeneous of second degree.  

The Lagrangian contains spin-1 kinetic terms proportional to 
\be
{\cal L}\propto i\Big( {\cal N}_{IJ}\,F_{\mu\nu}^{+I}F^{+\mu\nu J}\
-\ \bar{\cal N}_{IJ}\,F_{\mu\nu}^{-I} F^{-\mu\nu J} 
\Big)\,, \nonumber 
\ee
where $F_{\m\n}^{\pm I}$ are the (anti-)selfdual field strengths 
and $\cal N$ is proportional to the second derivative of the 
function $\bar F(\bar X)$. In addition there are moment couplings 
(to the fermions, or to certain background fields, to be 
discussed later), so that the field strengths $F^{\pm I}_{\m\n}$ 
couple linearly to tensors ${\cal O}^{\pm\m\n}_I$, whose form is 
left unspecified at the moment. Defining 
\be
G^+_{\mu\nu I}={\cal N}_{IJ}F^{+J}_{\mu\nu} + {\cal O}_{\mu\nu 
I}^+\,, \label{defG}
\ee
and the corresponding anti-selfdual tensor that follows from 
complex conjugation, the Bianchi identities and equations of motion
for the Abelian gauge fields take the form
\begin{equation}
\partial^\mu \big(F^{+} -F^{-}\big){}^I_{\m\n} =
\partial^\mu \big(G^+ -G^-\big){}_{\m\n I} =0\,.
\label{Maxwell}
\end{equation}
These are invariant under the transformation
\be
\pmatrix{F^{\pm I}_{\mu\nu}\cr  G^\pm_{\mu\nu I}\cr} 
\longrightarrow  \pmatrix{U&Z\cr W&V\cr} \pmatrix{F^{\pm 
I}_{\mu\nu}\cr  G^\pm_{\mu\nu I}\cr}\,,\label{FGdual}
\ee
where $U^I_{\,J}$, $V_I^{\,J}$, $W_{IJ}$ and $Z^{IJ}$ are 
constant real  $(n+1)\times(n+1)$ submatrices.

{}From (\ref{defG}) and (\ref{FGdual}) one derives 
that $\cal N$ must transform as
\begin{equation}
\tilde{\cal N}_{IJ} = (V_I{}^K {\cal N}_{KL}+ W_{IL} )\,
\big[(U+ Z{\cal N})^{-1}\big]^L{}_J  \,.\label{nchange}
\end{equation}
To ensure that $\cal N$ remains a symmetric tensor, at 
least in the generic case, the transformation  
(\ref{FGdual}) must be an element of $Sp(2n+2,{\bf R})$  
(we disregard a uniform scale transformation). 
Furthermore the tensor $\cal O$ must change according to
\be
\tilde{\cal O}_{\mu\nu I}^+ = {\cal O}_{\mu\nu J}^+  \,[(U+Z{\cal 
N})^{-1}]^J{}_I\,. \label{ochange}
\ee
The required change of $\cal N$ is induced by a change of the scalar 
fields, implied by
\be
\pmatrix{X^{I}\cr  F_{I}\cr} \longrightarrow  \pmatrix{\tilde 
X^I\cr\tilde F_I\cr}=
\pmatrix{U&Z\cr W&V\cr} \pmatrix{X^{I}\cr  F_I\cr}\,. 
\label{transX}
\ee
In this transformation we include a change of $F_I$. Because the 
transformation is symplectic, one can show that 
the new quantities $\tilde F_I$ can be written as  
the derivatives of a new function $\tilde F(\tilde X)$. 
The new but equivalent set of equations of motion one obtains by 
means of the symplectic transformation (properly extended to other 
fields), follows from the Lagrangian based on $\tilde F$. 

It is possible to integrate \eqn{transX} and one finds 
\bea
\tilde F(\tilde X) &\,=\,& F(X)-{\textstyle{1\over2}}X^I F_I(X) 
\nonumber \\
&& + {\textstyle{1\over2}} \big(U^{\rm T}W\big)_{IJ}X^I X^J 
\nonumber \\
&&+{\textstyle{1\over2}}
\big(U^{\rm T}V + W^{\rm T}  Z\big)_I{}^J X^IF_J\nonumber \\
&&+{\textstyle{1\over2}} \big(Z^{\rm T}V\big){}^{IJ}F_I F_J \,, 
\label{newfunction}
\eea
up to a constant and terms linear in the $\tilde X^I$. In the coupling to 
supergravity, where the function must be homogeneous of second 
degree, such terms are obviously excluded\footnote{%
   The terms linear in $\tilde X$ in \eqn{newfunction} are 
   associated with constant translations in $\tilde 
   F_I$ in addition to the symplectic rotation shown in 
   \eqn{transX}. Likewise one may introduce constant shifts in
   $\tilde X^I$. Henceforth we ignore these shifts, which are 
   excluded for local supersymmetry, even in the presence of a 
   background. Constant contributions to $F(X)$ are always 
   irrelevant.}. %

The above expression \eqn{newfunction} is not so useful, as it 
requires substituting $\tilde X^I$ in terms of $X^I$, or vice 
versa. When $F$ remains unchanged, $\tilde F(\tilde X) = F(\tilde 
X)$,  the theory is {\it invariant} under the 
corresponding transformations, but again this is hard to verify 
explicitly in this form. A more convenient method instead, is to verify 
that the substitution $X^ I\to \tilde X^I$ into the derivatives 
$F_I(X)$ correctly induces the symplectic transformations on the 
periods $(X^I,F_J)$. 

The result \eqn{newfunction} shows immediately that  
\be
F(X)-{\textstyle{1\over2}}X^I F_I(X) \label{holfunct}
\ee
transforms as a function under the symplectic transformations, 
i.e., as $\tilde f(\tilde X) = f(X)$, something that is obviously 
not true for the quantity $F(X)$. Here we should stress that, 
although we generically denote quantities such as $F(X)$ as  
holomorphic functions, we now introduce an important distinction 
by insisting that certain quantities transform as functions {\it under 
symplectic tranformations}. As we shall see, quantities with 
this property are more rare. 

In the coupling to supergravity, the above statement has no 
content, as \eqn{holfunct} vanishes identically by virtue of the 
homogeneity of $F(X)$. In other situations, however, the fact that 
\eqn{holfunct} transforms as a function under symplectic 
transformations is relevant. For example, the result has appeared in the 
literature in the context of the effective action of 
supersymmetric Yang-Mills theories, where it forms an 
ingredient in proving that \eqn{holfunct} must in fact 
be invariant under a certain subgroup of the symplectic 
transformations. To see how this goes, we first note that 
\be
\d \Big(2F(X)- X^I F_I(X) \Big) = (\d X^I) F_I - X^I(\d F_I) \,, 
\label{variation} 
\ee
under arbitrary variations. In particular we may consider changes 
of the moduli $u_\a$ that parametrize the ground states of the 
theory and identify $\d$ in the above equation with 
derivatives with respect to these moduli. When the $(X^I,F_J)$ 
can be defined  
in terms of the periods of a certain differential on a Riemann 
surface \cite{SW}, whose moduli space is that of the Yang-Mills 
theory, they are subject to Picard-Fuchs equations. The latter 
are partial differential equations involving multiple derivatives of the 
periods with respect to the $u_\a$. Combining these 
Picard-Fuchs equations with \eqn{variation}, one can show that 
$2F(X)- X^I F_I(X)$ satisfies a similar differential 
equation, which restricts it to a certain  
polynomial of the $u_\a$. For the details of this, we refer to 
\cite{matone,STY,EY}. 

What we like to stress here is the following. Knowing that 
\eqn{holfunct} can somehow be expressed in terms of the moduli, 
we conclude that it will not   
change when applying symplectic transformations belonging to 
the monodromy group. Hence, using that \eqn{holfunct} 
transforms as a {\it function} under symplectic transformations, 
we derive that it must be a function that is invariant under the 
monodromy subgroup.

\section{Backgrounds}
We now reconsider supersymmetric Yang-Mills theory in the 
presence of a chiral background field and a conformal 
supergravity background.  To couple supersymmetric vector multiplets 
to (scalar) chiral background fields is straightforward. One 
simply incorporates additional chiral fields $\Phi$ into the 
function $F$ that appears in the integrand of \eqn{chiral}, 
\be
S\propto {\rm Im}\;\Big(\int {\rm d}^4x \;{\rm d}^4\theta\; 
F(W^I,\Phi)\Big)\,.  \label{chiral2}
\ee
Also the coupling to conformal supergravity is
known \cite{DWLVP}. We draw attention to the fact that the $W^I$ 
are reduced, while the $\Phi$ can be either reduced or general 
chiral fields.  

Let us briefly discuss a few situations where such chiral 
backgrounds are relevant; in the next section we turn to explicit 
formulae. In supersymmetric theories many of the parameters 
(coupling constants, masses) can be regarded as  
background fields that are frozen to constant values (so that 
supersymmetry is left intact). Because these background fields 
correspond to certain representations of supersymmetry, the way 
in which they appear in the theory -- usually both perturbatively 
as well as nonperturbatively -- is restricted by supersymmetry. 
In this way we may derive restrictions on the way in which 
parameters can appear. An example is, for instance, the coupling 
constant and $\theta$ angle of a supersymmetric gauge theory, 
which can be regarded as a chiral field frozen to a complex 
constant $iS= \theta/2\pi+i4\pi/g^2$. 
Supersymmetry now requires that the function $F(X)$ depends on 
$S$, but {\it not} on its complex conjugate. This strategy of 
introducing so-called {\it spurion} fields is not new. In the 
context of supersymmetry it has been used in, for instance, 
\cite{shifman,amati,seiberg} to derive nonrenormalization 
theorems and even exact results. 
                      
Spurion fields can also be used for mass terms of 
hypermultiplets. When considering  
the effective action after integrating out the hypermultiplets, 
the dependence on these mass parameters can be incorporated in 
chiral background fields. In this example the background must 
be restricted to reduced chiral multiplets. In the previous 
example this restriction is optional. 
On the other hand, it may also be advantageous to not restrict 
the background fields to constant values, in order to introduce 
an explicit breaking of supersymmetry \cite{GG,AGD}. 

Another context where chiral backgrounds are relevant concerns 
the coupling to the Weyl multiplet, which involves interactions 
of vector multiplets to the square of the Riemann tensor. In this 
case the scalar chiral background  
is not reduced and is proportional to the square of the Weyl 
multiplet. Here the strategy is not, of course, to freeze the 
background to a constant, but one is interested in more general 
couplings with conformal supergravity. We return to 
a more detailed discussion of the Weyl multiplet shortly.                             

We add that all of this is very natural from the point 
of view of string theory, where the moduli fields, which 
characterize the parameters of the (supersymmetric) low-energy 
physics, reside in supermultiplets. In heterotic $N=2$ 
compactifications the 
background field $S$ introduced above coincides with the complex 
dilaton field, which comprises the dilaton and the axion, and 
belongs to a vector multiplet. The dilaton acts as the 
loop-counting parameter for string perturbation theory. Although 
the full supermultiplet that contains the dilaton is now 
physical, the derivation of nonrenormalization theorems can proceed
in the same way \cite{nilles,DWKLL}. We should stress here that 
when restricting the background to a reduced chiral multiplet, 
one can just treat it as an additional (albeit external) 
vector multiplet. Under these circumstances one may consider 
extensions of the symplectic  
transformations that involve also the background itself. Of 
course, when freezing the background to constant values, one must 
restrict the symplectic transformations accordingly. The above 
strategy is especially useful when dealing with anomalous 
symmetries. By extending anomalous transformations to the 
background fields, the  variation of these fields can compensate 
for the anomaly. The extended non-anomalous symmetry becomes
again anomalous once the background is frozen to a contant 
value. This strategy is particularly valuable when dealing with 
massive hypermultiplets. 

In the second half of this section we discuss a number of 
features pertaining to the Weyl multiplet \cite{BDRDW}. The bosonic 
fields of $N=2$ conformal supergravity consist of the vierbein 
field $e_\m^a$, $SU(2)\times U(1)$ chiral gauge fields $A_\m$ and 
$V^i_\m{}_j$ (associated with the automorphism group of the 
supersymmetry algebra), a selfdual tensor field $T_{abij}$, 
antisymmetric in both Lorentz and $SU(2)$ indices, and a scalar 
field $D$. The fermionic fields, which we will mostly ignore 
here, are the gravitino fields $\psi_\m^i$ and a spinor field 
$\chi^i$. 
In coupling the chiral Lagrangian \eqn{chiral2} to conformal 
supergravity, all of these fields will appear. However, the 
covariant quantities associated with this field representation 
form themselves a chiral selfdual tensor multiplet. Its lowest-weight 
component is the antisymmetric tensor  
$T^{ij}_{ab}$, the next one consists of $\chi^i$ and the 
field-strength of the gravitino field (with modifications such 
that it is superconformally covariant); then, at level 
$\theta^2$ we have the selfdual component of the Riemann tensor 
and the $SU(2)\times U(1)$ field strengths (all of them with 
proper superconformal modifications) as well as  
$D$ and $T_{abij}T^{ij}_{cd}$. All 
higher-$\theta$ components are dependent, in view of the  
fact that we are dealing with a reduced chiral multiplet. 

The square of the Weyl multiplet constitutes a scalar chiral 
field of scaling weight 2. Its lowest component is equal to 
$(\varepsilon_{ij} T^{ij}_{ab})^2$; the highest-$\theta$ component 
contains the square of the selfdual components of the Riemann 
tensor and the chiral field strengths, as well as a variety of 
other terms. We refer to \cite{BDRDW} for details. 

By associating the chiral background field $\Phi$ with the square 
of the Weyl multiplet $\cal W$, we thus obtain new (higher-derivative) 
couplings of vector multiplets with conformal supergravity. Assume 
that the function $F$ can be expanded as a power series,
\be
F(X, {\cal W}^2) = \sum_{g=0}^\infty \;F^{(g)}(X)\, \big({\cal 
W}^2\big)^g \,.\label{CYW2}
\ee
Because it must be homogeneous of second 
degree with scaling weights of $X$ and $\cal W$ that are both equal 
to unity, the coefficient functions $F^{(g)}(X)$ are homogeneous 
of degree $2(1-g)$. 
  
In supergravity the $X^I$ are not independent scalar fields, but 
are defined projectively; in more mathematical terms they can be 
regarded as sections of a complex line bundle. 
These sections can be expressed holomorphically in terms of 
independent complex fields $z^A$, which describe the physical 
scalars of the vector multiplets. The original quantities $X^I$ and 
the holomorphic sections $X^I(z)$ differ by a factor $m_{\rm 
P}\exp(K/2)$, where $K$ is the K\"ahler potential\footnote{%
   In terms of the holomorphic sections the K\"ahler potential 
   takes the form 
   $$
   K(z,\bar z)= -\log\big(i \bar X^I(\bar z)\,F_I(X(z)) -i X^I(z)\,
   \bar F_I(\bar X(\bar z))\big) \,.
   $$ 
   The K\"ahler metric is defined as $g_{A\bar B}= 
   \pa_A\pa_{\bar B} K(z,\bar z)$. Under projective 
   transformations of the 
   holomorphic sections, $X^I \to \exp (f(z)) \,X^I$, the K\"ahler 
   potential transforms by a K\"ahler transformation, so that the 
   metric remains invariant.} %
and $m_{\rm P}$ is the Planck mass. 
In view of the projective nature of the $X^I$, there 
is thus always one more physical vector field than there are physical 
scalars. The extra vector corresponds to the graviphoton.  
The Lagrangian encoded by \eqn{CYW2} gives rise to 
terms proportional to the square of the Riemann tensor times 
$(\varepsilon_{ij} T^{ij}_{ab})^{2(g-1)}$. After extracting the 
scale factor $m_{\rm P}\,\exp(K/2)$, the coefficient  
functions $F^{(g)}(X)$ give rise to holomorphic functions ${\cal 
F}^{(g)}$ of the $z$ (or rather sections of a line bundle).    

Let us consider the case where the function \eqn{CYW2} encodes 
the $N=2$ supersymmetric  
effective low-energy field theory corresponding to a type-II 
string compactification on a Calabi-Yau manifold. 
The Planck mass $m_{\rm P}$ is equal to the string scale 
divided by the string coupling constant $g_{\rm S}$; the latter 
is proportional to the dilaton. In a type-II string 
compactification the dilaton, which counts string loops, does not 
reside in the  
vector multiplet sector, so that the prefactor $\exp(K/2)$ is 
independent of the string coupling constant and the $X^I$ are 
proportional to $g_{\rm S}^{-1}$. Consequently  
$\big({\cal W}^2\big)^g$ is multiplied by terms of order $g_{\rm  
S}^{2(g-1)}$, which thus represent $g$-loop contributions in string 
perturbation theory. The coefficient functions can be 
determined in string theory from certain type-II string 
amplitudes  \cite{AGNT1} and indeed arise in the 
appropriate orders in string perturbation theory. An interesting 
feature is that the ${\cal F}^{(g)}$ can be 
identified with the topological partition function of a twisted 
nonlinear sigma model on a Calabi-Yau target space, defined on a 
two-dimensional base space equal to a genus-$g$ Riemann surface. 
The partition function is obtained by integrating 
appropriately over all these Riemann surfaces 
\cite{BCOV}. However, the partition functions 
${\cal F}^{(g)}$ do not depend holomorphically on the Calabi-Yau 
moduli. They  exhibit a so-called 
holomorphic anomaly due to the propagation of massless states, or 
equivalently, due to certain contributions from the boundary of the 
moduli space ${\cal M}_g$ associated with the genus-$g$ Riemann 
surfaces. The  holomorphic anomaly is governed by the following 
equations (with certain normalizations of the ${\cal F}^{(g)}$) 
\cite{BCOV},  
\bea
&&\pa_{\bar A} {\cal F}^{(g)} = \ft12 e^{2K} \bar {\cal W}_{\bar 
A}{}^{BC}\\ 
&&\qquad \times \Big[ D_BD_C{\cal F}^{(g-1)} + \sum _{r=1}^{g-1} 
D_B{\cal F}^{(r)}\,D_C {\cal F}^{(g-r)}\Big] \,,\label{anomalyeq} 
\nonumber 
\eea
for $g>1$, whereas for $g=1$  we have
\bea
\pa_A\pa _{\bar B} {\cal F}^{(1)} &=& \ft12 e^{2K}\, {\cal W}_{ACD} 
\bar{\cal W}_{\bar B}{}^{CD} -\ft1{24} \chi \, g_{A\bar B} 
\nonumber \\
&=&- \ft12 R_{A\bar B} + \ft12(n+1 -\ft1{12} \chi) g_{A\bar B}\,,
\label{F1}
\eea
with $R_{A\bar B}$ the Ricci tensor and  $\chi$ the Euler number  
of the Calabi-Yau moduli space.\footnote{%
   Here we used that the moduli space is an $n$-dimensional special 
   K\"ahler space. A particular solution of  
   \eqn{F1} is ${\cal F}^{(1)}=  -\ft12 \ln g +\ft12(n+1 
   -\ft1{12} \chi) K$, where $g$ is the metric determinant. } %
In these equations target-space 
indices are raised or lowered by means of the K\"ahler metric. 
Covariant derivatives are projectively covariant and defined by 
\be
D_A{\cal F}^{(g)} = \big(\pa_A +2 (1-g) \pa_A K\big) {\cal 
F}^{(g)} \,, 
\ee
and furthermore we used the definition 
\bea
&&{\cal W}_{ABC} =i  F_{IJK}\big(X(z)\big) \label{defW}\\
&&\qquad \qquad \quad\times {\partial X^I(z)\over \partial z^A}  
{\partial X^J(z)\over \partial z^B} 
{\partial X^K(z)\over \partial z^C} \,. \nonumber
\eea

In $N=2$ compatifications of the heterotic string the counting 
of string loops runs differently, because here the  
K\"ahler potential depends explicitly on the dilaton field. 
Now the dilaton dependence in $\exp(K/2)$ cancels  
the string coupling constant induced by the Planck mass,
so that the $X^I$ are generically of order zero in the string coupling 
constant. However, the dilaton coincides with one of the fields 
$z$ (or some function of them). The actual dependence on the 
string coupling constant is therefore directly governed by the 
explicit dilaton dependence of the quantities  
${\cal F}^{(g)}$. Interestingly enough, this explicit dependence 
is restricted, at least in perturbation theory, by virtue of a 
nonrenormalization theorem. Based on this theorem, in heterotic 
compactifications, one expects the ${\cal F}^{(g)}$ to appear 
at the one-loop level, with the   
exception of ${\cal F}^{(0)}$ and ${\cal F}^{(1)}$ which also 
receive classical contributions \cite{AGNT2}. Beyond  
this there are of course nonperturbative terms. 

The above observations are relevant in certain explicit tests of 
`string duality' between heterotic string compactifications on 
$K_3\times T_2$ and type-II string compactifications on  
Calabi-Yau manifolds \cite{KV}. In these tests 
\cite{AGNT2,KLT,KLM} the nonperturbative effects on the 
Calabi-Yau side are compared to the perturbative effects on the 
heterotic side. The latter were studied in \cite{DWKLL,AFGNT}.

\section{Duality in a chiral background}
After this somewhat qualitative discussion let us turn to more 
explicit results. In this section we verify the consistency 
of symplectic reparametrizations in a general chiral background. 
From this we learn how coefficient functions such as in 
\eqn{CYW2} change under these reparametrizations. As it 
turns out, they do not, in general, transform as 
functions, as is already suggested by the transformation rule 
\eqn{newfunction} for $F(X)$. 

The subsequent discussion is based on the action \eqn{chiral2}. 
A first observation is that, a priori, it is not  
meaningful to restrict the dependence on the background field. For 
instance, one may couple the theory linearly to the background, 
so that $\cal N$ will depend at most linearly on the background 
field. However, after a symplectic transformation, $\cal N$ will 
generically have a nonlinear dependence on the background, as 
follows from \eqn{nchange}. Therefore 
the only meaningful approach is to start from functions $F$ which 
depend both on the gauge superfield strengths $W^I$ and on the background 
field $\Phi$ in a way that is a priori unrestricted. Then one can 
proceed exactly as before and examine the equivalence classes in 
the presence of the background. The transformation rules, 
however, will also depend on the background fields. This does 
not affect the derivation, although there are a number of new 
features. 

We will consider the component expression corresponding to 
\eqn{chiral2} ignoring the fermions. We assume the presence of 
one chiral scalar background  
(the generalization to more background fields is 
straightforward), which itself may be equal to a scalar expression of 
nonscalar chiral fields as in \eqn{CYW2}. We denote the bosonic 
components of the chiral background superfield by   
$\hat A$, $\hat B_{ij}$, $\hat F^-_{ab}$ and $\hat C$. Here $\hat A$ 
and $\hat C$ are complex scalars, appearing at the $\theta^0$ 
and $\theta^4$ level of the chiral superfield, while the 
symmetric complex $SU(2)$ tensor $\hat B_{ij}$ and the 
anti-selfdual Lorentz tensor $\hat F^-_{ab}$ reside at the 
$\theta^2$ level. The holomorphic function $F$  now depends on 
the lowest-$\theta$ components $X$ and $\hat A$ of the gauge 
superfield strengths and the background field, respectively. The 
supersymmetric Lagrangian is proportional to the {\it real} part 
of the following expression
\bea
&&-2i F_I \,(\partial_\mu-iA_\mu)^2 \bar X^I   \nonumber  \\
&&-\ft14i  F_{IJ}\, Y^I_{ij} Y^{Jij} - \ft12 i \hat 
B_{ij}\,F_{AI}  Y^{Iij}   \nonumber\\
&&+\ft12 i F_{IJ} (F^{-I}_{ab} -\ft 14 \bar X^I 
T_{ab}^{ij}\varepsilon_{ij})(F^{-J}_{ab} -\ft14 \bar X^J 
T_{ab}^{ij}\varepsilon_{ij})  \nonumber\\
&&-\ft14 i F_I(F^{+I}_{ab} -\ft14  X^I 
T_{abij}\varepsilon^{ij}) T_{abij}\varepsilon^{ij}  \nonumber \\
&&-i \hat F^-_{ab}\, (F_{AI} F^{-I}_{ab} - \ft14 F_{AI} \bar X^I 
T_{ab}^{ij}\varepsilon_{ij})   \nonumber \\
&&+i F_A\hat C -\ft14 i F_{AA}(\varepsilon^{ik}\varepsilon^{jl}  \hat B_{ij} 
\hat B_{kl} -2 \hat F^-_{ab}\hat F^-_{ab})\cr 
&&-\ft1{16} iF (T_{abij}\varepsilon^{ij})^2- 
2i (\ft16 R-D) \,F_I \bar X^I\,, \label{basic}
\eea
where we also included the coupling to the bosonic fields of 
conformal supergravity. However, the vierbein determinant has 
been suppressed, while the $SU(2)$  
gauge fields do not appear because they couple only to fermions. 
Note that \eqn{basic} depends only on derivatives of the function 
$F(X,\hat A)$ with respect to the $ X^I$ and/or the background field 
$\hat A$. In the notation used previously we immediately 
derive the following expressions from \eqn{basic},\footnote{%
   Note that the expression for $\cal N$ takes this form 
   irrespective of whether we couple to supergravity. 
   This is so because the auxiliary tensor $T$ has not been 
   integrated out; therefore we must insist that the function 
   $F$ exists. After integrating out the auxiliary tensor it is 
   possible to reformulate the theory in such a way that the 
   function $F$ no longer needs to exist, as long as the periods 
   $(X^I,F_J)$ can consistently be written down \cite{Ceresole}.}%
\bea
{\cal N}_{IJ} &\,=\,&\bar F_{IJ}\,, \nonumber \\
{\cal O}_{\mu\nu I}^+    &=& \ft14 (F_I-\bar F_{IJ}X^J )T_{\mu\nu 
ij}\varepsilon^{ij} -\hat F^+_{\mu\nu} \,\bar  
F_{IA} \,. \label{NO} 
\eea

It is convenient to introduce the following definitions, 
\bea
{\partial\tilde X^I\over\partial X^J}\,&\equiv& \,{\cal 
S}^I{}_{\!J}(X,\hat A)
= U^I{}_{\!J} +Z^{IK}\,F_{KJ} \,, \nonumber\\
{\cal Z}^{IJ}\,&\equiv& \,[{\cal S}^{-1}]^I{}_K\, Z^{KJ}\,, \\ 
N_{IJ}\,&\equiv&\, 2 \,{\rm Im} \,F_{IJ}\,, \qquad N^{IJ}\equiv 
\big[N^{-1}\big]^{IJ}\,. \nonumber
\eea
The quantity ${\cal Z}^{IJ}$ is symmetric in $I$ and $J$, because 
$Z\,U^{\rm T}$ is a symmetric matrix as a consequence of the fact 
that $U$ and $Z$ are certain submatrices of a symplectic matrix. 

The symplectic reparametrizations should act on 
the quantities \eqn{NO} according to (\ref{nchange}) and 
(\ref{ochange}), which in the above notation read 
\bea
\tilde{\cal N}_{IJ} \,&=&\, (V_I{}^K {\cal N}_{KL}+ W_{IL} )\,
[\bar{\cal S}^{-1}]^L{}_J \,,\nonumber \\  
\tilde{\cal O}_{\mu\nu I}^+ \,&=&\, {\cal O}_{\mu\nu J}^+  \,
[\bar{\cal S}^{-1}]^J{}_I \,. \label{NOchange}
\eea
To ensure that these transformations are indeed realized, one 
introduces the transformations on the fields $X^I$, exactly as in 
section~2, except that the various quantities will now depend on 
the background field. So we have the transformation rule \eqn{transX} 
and the same expression \eqn{newfunction} for the new function 
after a symplectic transformation. Obviously the relation between $X$ and 
$\tilde X$ involves $\hat A$. 

Irrespective of the background the quantity $\cN$ still 
transforms according to the first equation of 
\eqn{NOchange}. Also the following result,
\bea
&&\tilde F(\tilde X,\hat A)-{\textstyle{1\over2}}\tilde X^I \tilde 
F_I(\tilde X,\hat A) =\nonumber \\
&&\qquad \qquad F(X,\hat A)-{\textstyle{1\over2}} X^I 
F_I(X,\hat A) \,,  \label{holfunct2}
\eea
still holds, so that there is a holomorphic function that 
transforms as a function under symplectic transformations. In the 
coupling to supergravity this result is still relevant, provided  
the background field $\hat A$ has a nonzero scaling weight. 
Other results which hold irrespective of the background, are 
\bea
\tilde N_{IJ} &=& N_{KL}\, \big[\bar{\cal 
S}^{-1}\big]^K{}_{\!I}\,  \big[{\cal S}^{-1}\big]^L{}_{\!J}\,, 
\nonumber\\ 
\tilde N^{IJ} &=& N^{KL}\, \bar{\cal 
S}^I{}_{\!K}\,{\cal S}^J{}_{\!L}\,, \\  
\tilde F_{IJK} &=& F_{MNP}\, \big[{\cal S}^{-1}\big]^M{}_{\!I} \,
    \big[{\cal S}^{-1}\big]^N{}_{\!J}\, \big[{\cal 
S}^{-1}\big]^P{}_{\!K} \,, \nonumber
\eea
where all quantities depend on both the fields $X$ and $\tilde 
A$. The symmetry of the first two quantities is preserved 
owing to the symplectic nature of the transformation. 

Results that specifically refer to the background are obtained 
by taking derivatives of $\tilde F$ (cf. \eqn{newfunction}), 
keeping $\tilde X$ fixed in partial differentiations of  
$\tilde F$ with respect to $\hat A$, and/or using already known 
transformations. In this way we obtain, for instance,
\bea
\tilde F_A(\tilde X,\hat A)&=&  F_A(X,\hat A)\,, \nonumber\\
\tilde F_{AI} &=& F_{AJ}\,[{\cal S}^{-1}]^J{}_I \,,\nonumber\\
\tilde F_I-\tilde F_{IJ}\tilde X^J  &=& [ F_J- F_{JK} X^K]\,
[{\cal S}^{-1}]^J{}_I\,, \label{vgl1}\\
\tilde F_I-\tilde{\bar F}_{IJ}\tilde X^J  &=& [F_J-{\bar F}_{JK} 
X^K]\,[\bar{\cal S}^{-1}]^J{}_I \,, \nonumber\\
\tilde F_{AA}(\tilde X,\hat A)&=& F_{AA}(X,\hat A) \nonumber\\
&&\quad -F_{AI}(X,\hat 
A)\,F_{AJ}(X,\hat A)\, {\cal Z}^{IJ}\,. \nonumber
\eea
These relations suffice to show that the transformation 
behaviour of the tensors ${\cal O}^\pm$ defined by \eqn{NO} 
is indeed in accord with \eqn{NOchange}. 

For later use we also list a few results involving higher 
derivatives of $F(X,\hat A)$,
\bea
\tilde F_{AIJ}&=& \big(F_{AKL} - F_{AM}{\cal Z}^{MN}F_{NKL}\big) 
\nonumber\\
&&\qquad \times
[{\cal S}^{-1}]^K{}_I \,[{\cal S}^{-1}]^L{}_J\,,  \nonumber\\
\tilde F_{AAI} &=& (F_{AAJ} -2 F_{AK}{\cal Z}^{KL} F_{ALJ} 
\nonumber \\
&& + F_{JKL}\, ({\cal Z}^{KM}F_{AM}) \,({\cal 
Z}^{LN}F_{AN})) [{\cal S}^{-1}]^J{}_I \,,\nonumber \\ 
\tilde F_{AAA}&= & F_{AAA} -3 F_{AAI}\,{\cal Z}^{IJ} F_{AJ}  
\label{vgl2}\\
&& + 3
F_{AIJ} \,({\cal Z}^{IK}F_{AK})({\cal Z}^{JL}F_{AL}) \nonumber\\
&& - F_{IJK}\,({\cal Z}^{IL}F_{AL})({\cal 
Z}^{JM}F_{AM})({\cal Z}^{KN}F_{AN})\,.   \nonumber 
\eea 

What remains to show is that the full equations of motion and the 
Bianchi identities are equivalent under symplectic 
reparametrizations. This can be done by identifying the terms in 
the Lagrangian that vanish by virtue 
of the equations of motion for the auxiliary fields and the 
vector fields. For the remaining terms one must then prove 
that they preserve their form under symplectic transformations. 
To do this we write the Lagrangian as follows,
\bea
{\cal L}&=& \ft14 N_{IJ}\,{\cal Y}_{ij}^I {\cal Y}^{ijJ} 
\nonumber\\
&&-\ft12 i(\partial_\mu W_\nu^I-\partial_\nu W_\mu^I) 
(G^+ -G^-)^{\mu\nu}_I\nonumber \\
& &-2i F_I \,(\partial_\mu-iA_\mu)^2 \bar X^I  + {\rm h.c.}  
\nonumber\\
&&+\ft12 \hat B_{ij}\hat B^{ij} \,N^{IJ} 
\bar F_{AI}F_{AJ} \nonumber   \\
&&+\ft14 [\hat B_{ij}\hat B_{kl} 
\varepsilon^{ik}\varepsilon^{jl} -2\hat F^-_{ab} \hat F^-_{ab}] \nonumber\\
 &&\quad\quad  \times [ N^{IJ} F_{AI} F_{AJ} -iF_{AA}] + {\rm 
h.c.}  \label{lagrangian}\\
&&-\ft18 i (G^-_{abI}\bar X^I -F^{-I}_{ab}\bar F_I) 
T_{ab}^{ij}\varepsilon_{ij}+{\rm h.c.} \nonumber\\
&&+\ft12 \hat F_{ab}^-( F^{-K}_{ab} \bar F_{KI} - 
G^-_{abI}) N^{IJ} F_{AJ} +
{\rm h.c.} \nonumber\\
&&+iF_A\hat C + {\rm h.c.}  \nonumber\\
&&+\ft1{16} i  (\bar F-\ft12 \bar X^I \bar F_I ) 
(T_{ab}^{ij}\varepsilon_{ij})^2 + {\rm h.c.} \nonumber\\
&&-\ft1{8}  \hat F_{ab}^- T_{ab}^{ij}\varepsilon_{ij} 
(\bar F_{IJ}\bar X^J - \bar F_I )N^{IK} F_{AK}+ 
{\rm h.c.}  \nonumber\\
&&-2 i (\ft16 R-D) (\bar X ^IF_I -\bar F_IX^I)\,,\nonumber
\eea
where we redefined the auxiliary fields $Y^I_{ij}$ according to  
$$
{\cal Y}^I_{ij} = Y^I_{ij}- iN^{IJ} 
[\hat B_{ij} F_{AJ} - \varepsilon_{ik}\varepsilon_{jl}  \hat 
B^{kl} \bar F_{AJ}] \,.
$$
The field equation of the auxiliary fields puts $\cal Y$ to zero, 
so that we can drop the first term in \eqn{lagrangian}. Likewise, 
the field equations for the vector fields converts the second term in 
\eqn{lagrangian} into a total divergence. 

Using the identities (\ref{holfunct2}-\ref{vgl1}) it 
is not difficult to show that all the terms in \eqn{lagrangian}, 
with the exception of the first two, preserve their form under 
symplectic transformations.\footnote{%
    Here we note a useful theorem \cite{DWVVP}: from a symplectic 
    vector, such as $(F^{\pm I}_{\m\n}, G^\pm_{\m\n}{}_J)$, one can 
    construct a quantity ${\cal V}_I= (G^\pm_{\m\n}{}_I-F_{IJ} 
    F^{\pm J}_{\m\n})$ which transforms under symplectic 
   transformations as $\tilde {\cal V}_I= {\cal V}_J\, [{\cal 
   S}^{-1}]^J{}_I$. The third and fourth equation of \eqn{vgl1} 
   can be derived directly on the basis of this theorem.} %
This concludes the proof that duality transformations are 
consistent in the presence of a chiral background with coupling 
to supergravity.

We close this section by pointing out that restricting the 
background to a constant (i.e., $\hat A$ constant and all other 
background components vanishing) just gives the standard coupling 
of vector  
multiplets to $N=2$ conformal supergravity. On the other hand, 
keeping also the fields $\hat B$ and/or $\hat C$ nonzero causes a 
breakdown of supersymmetry. This feature of the  
background field can be exploited when studying certain $N=1$ 
supersymmetric, or even nonsupersymmetric, theories.

\section{Symplectic functions and holomorphy}
The above results show that certain expressions can be 
constructed from the function $F(X,\hat A)$ and its derivatives 
that transform as {\it functions} under symplectic 
transformations. One obvious example  
is the holomorphic expression \eqn{holfunct};  another one 
is $F_A$, the first derivative of $F$ with respect to the background. 
However, higher than first derivatives of $F$ with respect to $\hat A$ 
do not transform as functions under symplectic transformations. 
This means that the coefficient functions in an expansion such as 
\eqn{CYW2} do not transform as symplectic functions, with the 
exception of ${\cal F}^{(1)}$. The transformation rules for these 
coefficient functions follow from \eqn{vgl1} and \eqn{vgl2} 
and their generalizations, putting the background field 
$\hat A$ to zero. 

This conclusion may be somewhat disturbing especially when considering 
symplectic transformations that constitute an invariance. In that 
situation we have $\tilde F(\tilde X,\hat A) = F(\tilde X,\hat A)$. 
In spite of that, this does not imply that the coefficient 
functions (i.e.  multiple derivatives with respect to the 
background) are invariant {\it functions} under the corresponding 
tranformations. This should only be the case for the first one 
corresponding to $F_A$.  

One may wonder whether there are modifications of the 
multiple-$\hat A$ derivatives of $F$ that do transform as 
functions under symplectic transformations. Such functions should 
be expected to arise when evaluating the coefficient functions 
directly on the basis of 
some underlying theory, such as string theory. These 
modifications seem possible in view of the fact that the combination
$$
F_{AA} + i N^{IJ} F_{AI} F_{AJ} \,,
$$
which appears in \eqn{lagrangian}, does indeed transform as a function 
under symplectic transformations. The latter was in fact required 
in order for the duality transformations to be consistent in the 
presence of the full chiral background, as shown in the previous 
section. Likewise, one may verify by explicit calcculation that 
there is a generalization of the third  derivative, 
\bea
F_{AAA} &&+ 3 i N^{IJ}F_{AAI}F_{AJ} \nonumber\\
&&  - 3 N^{IK} \,N^{JL}\,F_{AIJ}F_{AK}F_{AL}\nonumber\\  
&& - i N^{IL}\,N^{JM}\, N^{KN} \,F_{IJK} 
F_{AL}F_{AM}F_{AN}\,,\nonumber 
\eea
which also transforms as a symplectic function. 

It turns out that these functions can be generated 
systematically. Assume that $G(X,\hat A)$ transforms as a 
function under symplectic transformations. Then one readily 
proves that also ${\cal D}G(X,\hat A)$ transforms as a symplectic 
function, where\footnote{%
   We note that ${\cal D}$ and $N^{IJ}\pa_J$ commute. } %
\be 
{\cal D}\,\equiv\, {\pa\over \pa \hat A} +
iF_{AI}N^{IJ}{\pa\over \pa X^J} \,. 
\ee
Consequently one can directly 
write down a hierarchy of functions which are modifications of 
multiple derivatives $F_{A\cdots A}$,
\be
F^{(n)} (X,\hat A) \,\equiv \, {1\over n!}{\cal D}^{n-1} F_A(X,
\hat A)\, ,\label{Fn}
\ee
where we included a normalization factor. 
All the $F^{(n)}$ transform as functions under symplectic 
functions. However, except for the first one, they  
are not holomorphic. The lack of holomorphy is governed 
by the following equation ($n>1$),
\be
{\pa F^{(n)} \over \pa \bar X^I}= \ft12 \bar F_I{}^{JK} 
\sum_{r=1}^{n-1}\; {\pa F^{(r)} \over \pa  X^J}\,{\pa F^{(n-r)} 
\over \pa  X^K}\,, \label{anomalyeq2}
\ee
where $\bar F_I{}^{JK}= \bar F_{ILM}\,N^{LJ}N^{MK}$. 
Interestingly enough, this equation coincides with 
\eqn{anomalyeq}, except that the first term on the right-hand 
side of \eqn{anomalyeq} is absent here. This is the term that 
arises from Riemann surfaces where a closed loop is pinched, 
which lowers the genus by one unit. The 
second term, which coincides with the one above, corresponds to  
pinchings that separate the Riemann surface into two 
disconnected surfaces \cite{BCOV}. 

We should stress that \eqn{anomalyeq2} was obtained in a 
very general context and applies to both rigid and local $N=2$ 
supersymmetry. In the latter case we have to convert to 
holomorphic sections $X^I(z)$. This requires to set $\hat A=0$ in 
\eqn{anomalyeq2}. The holomorphic anomaly can thus be viewed as 
arising from a conflict between the requirements of holomorphy 
and of a proper behaviour under symplectic transformations. The 
fact that there is not an additional term in the anomaly 
equation, as in (\ref{anomalyeq}), is not unrelated to the fact that 
$F^{(1)}$ is still holomorphic.

\vspace{5mm}
\par
\noindent{ \bf Acknowledgements}\vspace{0.3cm}
\par
I am grateful for stimulating discussions with L. Alvarez-Gaum\'e, 
R. Dijkgraaf, A. Klemm, J. Louis, D. L\"ust, S. Theisen and 
E. Verlinde. 



\begin{thebibliography}{99}
\bibitem{SG} S. Ferrara, J. Scherk and B. Zumino, Nucl. Phys. {\bf B121}
(1977) 393.
\bibitem{DWVP}  B. de Wit and A. Van Proeyen, Nucl. Phys. {\bf
B245} (1984) 89.
\bibitem{CecFerGir} S. Cecotti, S. Ferrara and L. Girardello,
Int. J. Mod. Phys. {\bf A4} (1989) 2457.
\bibitem{SW}
N.~Seiberg and E.~Witten, Nucl. Phys. {\bf B426} (1994) 19; 
Erratum {\bf B430} (1994) 485; {\bf B431} (1994) 484.
\bibitem{Lerche} A. Klemm, W. Lerche, S. Theisen and S. 
Yankielowicz, Phys. Lett. {\bf B344} (1995) 169.\\
P. Argyres and A. Faraggi, Phys. Rev. Lett. {\bf 73} (1995) 3931. 
\bibitem{matone} M. Matone, Phys. Lett. {\bf B357} (1995) 342.
\bibitem{STY} J.~Sonnenschein, S. Theisen and S. Yank\-ielowicz, 
{\it On the relation between the holomorphic prepotential and the 
quantum moduli in SUSY gauge theories}, preprint TAUP-2295-95, 
LMU-TPW-95-15, hep-th/9510129.
\bibitem{EY} T. Eguchi and S-K Yang, {\it Prepotentials of $N=2$ 
supersymmetric gauge theories and soliton equations}, preprint 
UT-728, hep-th/9510183. 
\bibitem{DWLVP} B. de Wit, P.G. Lauwers and A. Van Proeyen, Nucl. Phys. {\bf
B255} (1985) 569.
\bibitem{shifman} M.A. Shifman and A.I. Vainshtein, Nucl. Phys. 
{\bf B277} (1986) 456; {\bf B359} (1991) 571. 
\bibitem{amati} D. Amati, K. Konishi, Y. Meurice, G.C. Rossi and 
G. Veneziano, Phys. Rep. {\bf 162} (1988) 169.
\bibitem{seiberg} K. Intrilligator, R.G. Leigh and N. Seiberg, 
Phys. Rev. {\bf D50} (1994) 1092.
\bibitem{GG} L. Girardello and M.T. Grisaru, Nucl. Phys. {\bf 
B194} (1982) 65.
\bibitem{AGD} L. Alvarez-Gaum\'e and J. Distler, in preparation. 
\bibitem{nilles} H.P. Nilles, Phys. Lett. {\bf 180B} (1986) 240.
\bibitem{DWKLL}  B. de Wit, V. Kaplunovsky,
J. Louis and D. L\"ust, Nucl. Phys. {\bf B451} (1995) 53.
\bibitem{BDRDW} E.A. Bergshoeff, M. de Roo and B. de Wit, Nucl. 
Phys. {\bf B182} (1981) 173.
\bibitem{AGNT1} I. Antoniadis, E.Gava, K. Narain and T.R. Taylor, 
Nucl. Phys. {\bf B413} (1994) 162.
\bibitem{BCOV} M. Bershadsky, S. Cecotti, H. Ooguri and C. Vafa, 
Nucl. Phys. {\bf B405} (1993) 279; Commun. Math. Phys. {\bf 165} 
(1994) 311.
\bibitem{AGNT2} I. Antoniadis, E.Gava, K. Narain and T.R. Taylor, 
Nucl. Phys. {\bf B455} (1995) 109.   
\bibitem{KV} S. Kachru and C. Vafa, Nucl. Phys. {\bf B450} (1995) 
69.
\bibitem{KLT} V. Kaplunovsky, J. Louis and S. Theisen, 
Phys. Lett. {\bf 357B} (1995) 71.
\bibitem{KLM}
A. Klemm, W. Lerche and P. Mayr, Phys. Lett. {\bf 357B} 
(1995) 313.
\bibitem{Ceresole} A. Ceresole, R. D'Auria, S. Ferrara and A. Van
Proeyen, Nucl. Phys. {\bf B444} (1995) 92.
\bibitem{AFGNT} I. Antoniadis, S. Ferrara, E. Gava, K.S. Narain and 
T.R. Taylor, {\em Nucl.~Phys.} {\bf B447} (1995) 35. 
\bibitem{DWVVP} B. de Wit, F. Vanderseypen and A. Van Proeyen,
Nucl. Phys. {\bf B400} (1993) 463, appendix~C.

\end{thebibliography}
\end{document}